\documentstyle[preprint,aps]{revtex}

\begin{document}
\draft
\preprint{}

\newcommand {\be}{\begin{equation}}
\newcommand {\ee}{\end{equation}}
\newcommand {\bea}{\begin{eqnarray}}
\newcommand {\eea}{\end{eqnarray}}
\newcommand {\nn}{\nonumber}

%
%

\title{ Photoemission Spectra in t-J Ladders with Two Legs}        

\author{Stephan Haas and Elbio Dagotto}

\address{Theoretische Physik,
ETH-Hoenggerberg, CH-8093 Zurich, Switzerland, \\
and
National High Magnetic Field Laboratory,
Florida State University, Tallahassee, FL 32306}

\date{\today}
\maketitle

\begin{abstract}

Photoemission spectra for the isotropic two-leg t-J ladder are calculated 
at various hole-doping levels using
Exact Diagonalization techniques. Low-energy sharp features
caused by short-range
antiferromagnetic correlations are observed at finite doping levels
close to half-filling, above the naive
Fermi momentum. These features should be observable in angle-resolved
photoemission experiments. In addition, the formation of a d-wave pairing
condensate as the ratio J/t is increased leads to dynamically generated
spectral weight for momenta close to ${\rm k_F}$ where the ${\rm
d_{x^2-y^2} }$-order parameter is large.

\end{abstract}


\narrowtext

%
%
The recent discovery of two-leg spin-1/2 ladders such as vanadyl
phosphorate ${\rm (VO)_2P_2O_7}$ and ${\rm SrCu_2O_3 }$ has received
considerable attention.\cite{johnston,azuma} On one hand they represent a
physical realization of RVB spin-liquids,\cite{review}
with a ground state dominated by spin singlets, mostly along the rungs.
On the other hand,
they are a lower-dimensional analogue of the much-studied
two-dimensional high-${\rm T_c}$ cuprates and thus a test-ground for
generic ideas underlying the physics of strongly correlated electrons.
Furthermore, recently progress has been made in doping ${\rm SrCu_2O_3
}$ with holes by substituting some of the Sr-sites by La.\cite{hiroi}
Then it is
of much interest to test predictions for the Fermi surface of
correlated
electrons in this novel geometry and contrast them with results in two
dimensions. 

In this paper we study the isotropic t-J model on a 
two-leg ladder defined by the Hamiltonian :
\bea
 {\rm H}&=&{\rm -t\sum_{j,\sigma} [ \sum_{a=1}^2  \tilde{c}^{\dagger}_{a\sigma}(j)
\tilde{c}_{a\sigma}(j+1) + \tilde{c}^{\dagger}_{1\sigma}(j)
\tilde{c}_{2\sigma}(j) +h.c.]  }\nonumber \\
  &+&{\rm J\sum_{j,a}[{\bf S}_a(j)\cdot{\bf S}_a(j+1)-{1\over4} n_a(j)n_a(j+1)]}\nonumber \\
  &+&{\rm J\sum_{j}[{\bf S}_1(j)\cdot{\bf S}_2(j)-{1\over4} n_1(j)n_2(j)],
}
\eea
where j is the rung index, and ${\sigma (=\uparrow,\downarrow) }$ and
a (=1,2) are spin and leg indices. A powerful way to study the Fermi
surface of this system is by Angle Resolved Photoemission
Spectroscopy (ARPES) which measures the imaginary part of the
one-hole Green's function ${\rm A_h({\bf p},\omega)}$, and
corresponds to the sudden removal of an electron from the material
(likewise the ${\rm A_e({\bf p},\omega)}$ corresponds to the process
of creating an additional electron). 

As shown in previous reports, two-leg ladders close to half-filling
belong to the universality class of Luther-Emery Liquids.\cite{tsunetsugu}
Then, it
is reasonable to consider the non-interacting picture of two coupled
chains to understand the topology of the Fermi surface for this
special geometry. In Fig. 1 the available single-particle states in
two coupled 8-site chains with periodic boundary conditions in the
chain-directions are shown for different filling levels. While at
finite hole doping a splitting between a bonding and an antibonding
band (of order J) is expected, at half-filling this splitting does not
occur since the kinetic energy is suppressed due to the constraint of
no double-occupancy.
In this case the bonding and the antibonding bond lie on
top of each other (indicated by the thick solid line), and the
single-particle states below the Fermi-momentum are shown as solid
circles while the unoccupied states are denoted by open circles. 
In this case the highest occupied level is completely filled, and
this constitutes a closed shell configuration. Upon doping with
two holes a splitting into a bonding and an antibonding band
occurs. Also, the highest occupied energy levels which
belong to the antibonding band (at
momenta ${\rm (\pi/4,\pi) }$ and ${\rm (-\pi/4,\pi) }$) are only
half-filled. Then, we have an open shell for this case. Finally, for 4
holes only the lowest state in the antibonding band is filled, while
the bonding band is half-filled for all of the above cases.


In the following, it will become apparent that the presence of
exchange interactions
modifies this simple picture by introducing additional features into the
naively expected electronic occupation structure predicted by the
non-interacting limit. For example, the presence of strong
antiferromagnetic (AF) correlations in the system favors the formation of
a magnetic superstructure, i.e. ``shadow bands" which were first
suggested by Kampf and Schrieffer in the context of the
two-dimensional cuprates close to half-filling.\cite{kampf} In subsequent
numerical studies it has been shown that an antiferromagnetic
correlation length of about 2-3 lattice spacings - as it is realized
e.g. in Bi2212 \cite{imai} - is sufficient to produce a shadow Fermi
surface that is barely resolvable by current PES techniques.\cite{haas}
On the other
hand, at half-filling and T=0 there is long-range antiferromagnetic order
in two dimensions, and thus a strong shadow signal is expected for
this case, as it was observed for ${\rm Sr_2CuO_2Cl_2}$.
\cite{wells}  In the two-leg ladder the situation differs slightly since
only short-range order with a correlation length of roughly 3.19
lattice spacings is found at half-filling.\cite{white}

Nevertheless, as is shown in Fig. 2, there is a clear shadow signal
for momenta ${\rm |k| > |k_F| (k_F = (\pi/2,0)) }$ in the bonding band
of 2-leg ladders.
In analogy to the two-dimensional case, a sharp peak at the lower
end of the PES is observed. It disperses with a bandwidth of order J
as can be seen by varying the ratio J/t. Also, this feature
becomes sharper as J/t is increased which implies that it is associated
mainly with the spin degrees of freedom in the system. Furthermore a
broader high-energy band located at binding energies of order t
is observed, corresponding
to incoherent excitations of the hole. As J/t is
increased much of the weight from the incoherent band is transferred
to the low-energy band. ARPES experiments in undoped 2-leg ladders
should be able to observe considerable weight above the naive Fermi 
momentum, as shown in Fig. 2.


The introduction of a hole into the half-filled system leads to the splitting
of the final states in ${\rm A_h({\bf p},\omega)}$ 
into a bonding and an antibonding band.
The hole goes into the antibonding band distorting the staggered
spin order of the half-filled system. Thus, shadow band features are
more visible in the bonding band especially at large values of J/t.

Upon doping the system with two holes, a main Fermi surface consistent
with the occupied sites indicated in Fig. 1 is observed. A
Fermi surface crossing of a sharp band dispersing as ${\rm -2
t_{eff} cos(k)}$ can be seen. The effective hopping ${\rm
t_{eff}}$ is smaller than t, as observed also in two-dimensional
clusters.\cite{moreo} However, as the doping level is increased
${\rm t_{eff} \rightarrow t}$. 
The Fermi momenta lie between ${\rm (\pi/2,0) }$ and ${\rm (3\pi/4,0)}$
for the bonding band and at ${\rm (\pi/4,\pi) }$ 
for the antibonding band.

In the bonding band, a sharp peak is observed at low binding energy
at momenta $(3\pi/4,0)$ and $(\pi,0)$, very similar to the shadow features
found at half-filling in Fig. 2. Note that by  
increasing the ratio J/t, an interesting feature occurs at momentum
$(\pi/2,\pi)$ in the PES, just above the antibonding ${\rm \bf k_F}$ 
for this filling : while there is little coherent
low-energy spectral
weight at this momentum for J/t=0.5, a sharp peak emerges
as J/t is increased. It is hence dynamically generated by
correlations that increase with J/t. As suggested recently by Hayward et
al. a ${\rm d_{x^2-y^2} }$ resonant valence bond
(RVB) phase may become stable in this parameter
regime.\cite{hayward,dagotto}  
We thus tentatively associate the PES weight which appears at
$(\pi/2,\pi)$ with increased ${\rm d_{x^2-y^2} }$ RVB correlations.
It may correspond to a Bogoliubov quasiparticle excitation made of
an electron and a hole (${\rm \alpha_{\bf k} = u_{\bf k} 
c^{\dagger}_{\bf k \uparrow} + v_{\bf k} c_{-{\bf k \downarrow}} }$).
\cite{tsunetsugu}


In principle, all holes and electrons close to ${\rm \bf k_F}$ can contribute 
to the formation of the
superconducting (SC)
condensate. Then, the following question arises : how much of the spectral
weight seen in the $(3\pi/4,0)$ and ${\rm (\pi,0)}$ PES peak 
is due to pairing and how much is due
to short-range AF correlations along the chains ?
To distinguish these two contributions 
to the spectral weight at 
${\rm (3\pi/4,0)}$ and ${\rm (\pi,0)}$
we include into the Hamiltonian (Eq. 1)
a nearest-neighbor interaction
term along the rungs, ${\rm V \sum_j n_1(j) n_2(j) }$, which 
disfavors a bound state of two holes on a rung if V is repulsive. For
either sign of V, the ${\rm d_{x^2-y^2} }$ character of the
bound state is destroyed if
the magnitude of V is chosen to be large enough. From an explicit
calculation of the spin-spin 
correlation functions (shown in Fig. 4)
we observe that in the presence of the nearest 
neighbor interaction term AF correlations are increased in the
chains when V is negative and decreased for positive V. 
This is consistent with the observation that the two-hole rung state is 
destroyed for a large enough positive V, spreading the second hole into the 
chains and thus smearing out the background staggered short-range
spin order at short distances.
However, the addition of a repulsive V-term with V as large as 5t only 
reduces the AF correlation length from 3 (at V=0) to about 2 lattice
spacings. Then, the main effect of the rung density-density repulsion is
to destroy the d-wave bound state, not altering much the staggered
spin-spin correlations. This is important for our discussion below on
the interpretation of some of the PES features observed in our
calculations.


In Fig. 5 PES are shown for a two-hole ground state at fixed J/t=1.0
for various values of V/t. Indeed the peak located at ${\rm (\pi/2,\pi)}$
for V/t=0 
disappears when the magnitude of V/t is large, indicating the suppression of
pairing. However, the peak at ${\rm (3\pi/4,0)}$ increases
for negative V and decreases for positive V having thus the same qualitative
dependence on V/t as the AF correlation length of the system.
Certainly it cannot be ruled out that there are small contributions 
from the now very weakly bound 2-hole state to the
low-energy spectral weight in the bonding band above ${\rm \bf k_F}$.
However, our calculation suggests that the 
superstructure observed in the bonding band is dominated by
backscattering processes characteristic of AF correlations.
Note that at V=-5t the bonding band is shifted away from the chemical
potential by a substantial amount reflecting the fact that an attractive
interaction across the rungs increases the band splitting.


The strength of backscattering processes in the bonding band
(with characteristic momentum ${\rm {\bf Q}_b = (\pi,0) }$)
and in the antibonding band (with characteristic momentum 
${\rm {\bf Q}_a = (\pi,\pi) }$) is reflected in the relative
spectral
intensities of the low-energy peak in ${\rm A_h({\bf Q}_b,\omega)}$
and ${\rm A_h({\bf Q}_a,\omega)}$ respectively.
While the coherent part in ${\rm A_h({\bf Q}_a,\omega)}$ is
reduced with negative V/t, it is enhanced in
${\rm A_h({\bf Q}_b,\omega)}$ and vice versa.

Before turning to higher hole-filling levels let us discuss the
Inverse Photoemission Spectra (IPES) shown in Fig. 3. While
for small values of J/t a rather coherent IPES band is observed,
for J/t=2.0 higher energy excitations carry considerable 
spectral weight. To understand this behavior we should remember
that a hole quasiparticle is made out of an actual hole (empty site)
plus a surrounding region where the AF correlations have been altered
with respect to those in the absence of carriers. This effect reduces
drastically the quasiparticle weight Z in the PES.\cite{schrieffer}
When a hole is annihilated in an IPES process
the spin-excitations of the ``dressed" quasiparticle are left
behind and form a low-energy
band of width J above ${\rm \bf k_F}$.\cite{eder} 
However, there are a variable number of such spin excitations,
i.e. multi-magnon processes which become more dominant at larger
J/t. Thus, more high-energy spectral weight is observed in the
IPES with increasing J/t.

Finally let us turn to a higher filling level, i.e. 4 holes.
The corresponding PES and IPES are shown in Fig. 6. The 
Fermi momentum in the antibonding band is  
now shifted with respect to the 2-hole ground state
according to Fig. 1. It lies between
${\rm (\pi/4,\pi)}$ and ${\rm (\pi/2,\pi)}$.
A shadow peak is observed in the bonding band at ${\rm 
(3\pi/4,0) }$ and ${\rm (\pi,0)}$
although considerably weaker than for the case of
2 holes. This corresponds to a shorter AF correlation length
for this doping. A feature possibly induced by d-wave pairing
correlations which sharpens rapidly
with increasing J/t appears at momentum
${\rm (\pi/4,\pi)}$. An increase in low-energy coherence is
also observed at ${\rm (\pi/2,\pi)}$.


The observation of robust antiferromagnetically induced sharp
peaks in two-leg ladder systems is not entirely surprising.
Remember that in these systems there is a robust spin-gap
that quite likely survives the introduction of a small density of
carriers.\cite{review} This spin-gap is caused by the tendency of spins
to form rung singlets, which favors hole pair formation in these
rungs. Once such pairs are formed, the staggered spin background 
remains almost intact.\cite{mila}

In conclusion, we have studied the Photoemission Spectra for two-leg
t-J ladders at various hole-doping levels. In analogy to the
two-dimensional case, sharp peaks induced by remnant short-ranged
antiferromagnetic correlations are
observed for small hole-doping levels.
Such robust peaks should be detectable in ARPES experiments applied to
ladder systems. Previous work has also indicated that similar
structure clearly appears in undoped spin-1/2 Heisenberg chains
even in the presence of spin-Peierls dimerization.\cite{cugeo}
Then, it is not necessary to have AF long-range order to observe ``shadow
bands". A mild AF correlation length of only 2 or 3 lattice spacings is 
enough, since the effect is dominated by short-range AF fluctuations.
In addition, PES weight for momenta close to ${\rm k_F}$ is
generated by increasing the ratio J/t. This dynamically generated
weight may signal the onset of
${\rm
d_{x^2-y^2} }$-order.

\medskip
We wish to thank T.M. Rice, T. M\"uller, D. Duffy, M. Sigrist and A. Moreo
for useful discussions.
E. D. is supported by grant NSF-DMR-9520776.
Additional support of the
the National High Magnetic Field Laboratory (NHMFL)
is acknowledged.

%
%

{\bf Figure Captions}

\begin{enumerate}

\item Schematic plot of the occupied (solid circles) and unoccupied (open
circles) states in a 2$\times$8 ladder system with periodic boundary
conditions.

\item 
Photoemission Spectra for a half-filled 2$\times$8 t-J ladder with periodic
boundary conditions. The $\delta$-functions have been given a finite width
of ${\rm \epsilon = 0.25t}$.

\item
Photoemission Spectra (solid lines) and
Inverse Photoemission Spectra (dashed lines)
for a 2$\times$8 t-J ladder with 2 holes. The position of the chemical
potential is indicated by the thin solid line.

\item
Hole-hole and spin-spin correlations
for a 2$\times$8 t-J-V ladder with 2 holes at J/t=1.0. The sites
on the first chain are labeled with j=1,...,8, and the sites on
the second chain have j=9,...,16.

\item
Photoemission Spectra
for a 2$\times$8 t-J-V ladder with 2 holes at J/t=1.0.

\item
Photoemission Spectra (solid lines) and
Inverse Photoemission Spectra (dashed lines)
for a 2$\times$8 t-J ladder with 4 holes.

\end{enumerate}

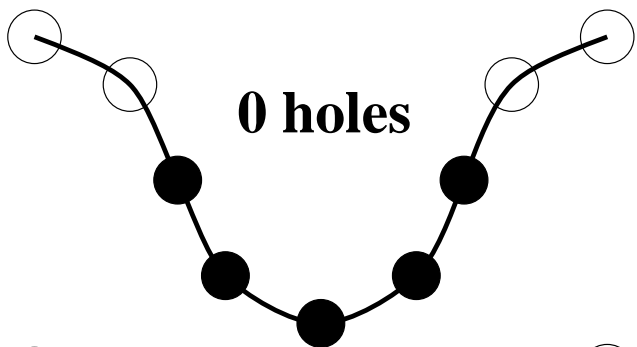
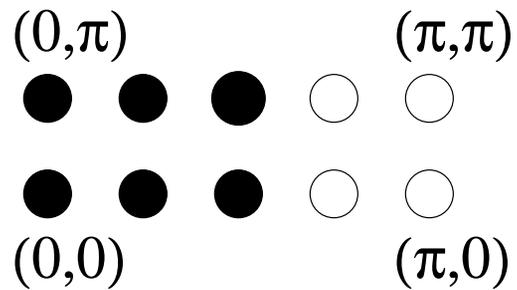
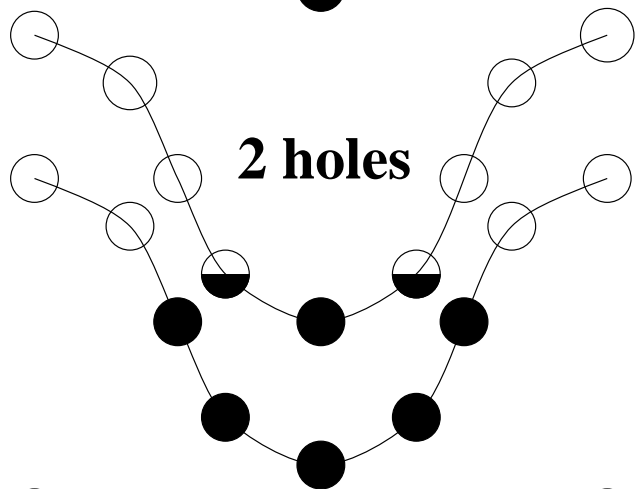
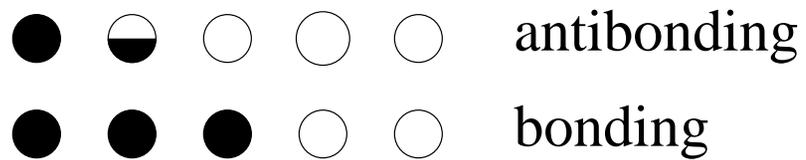
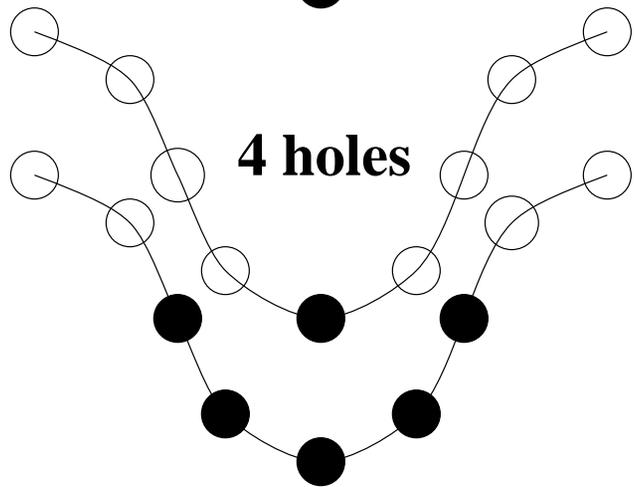
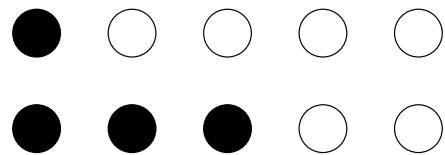
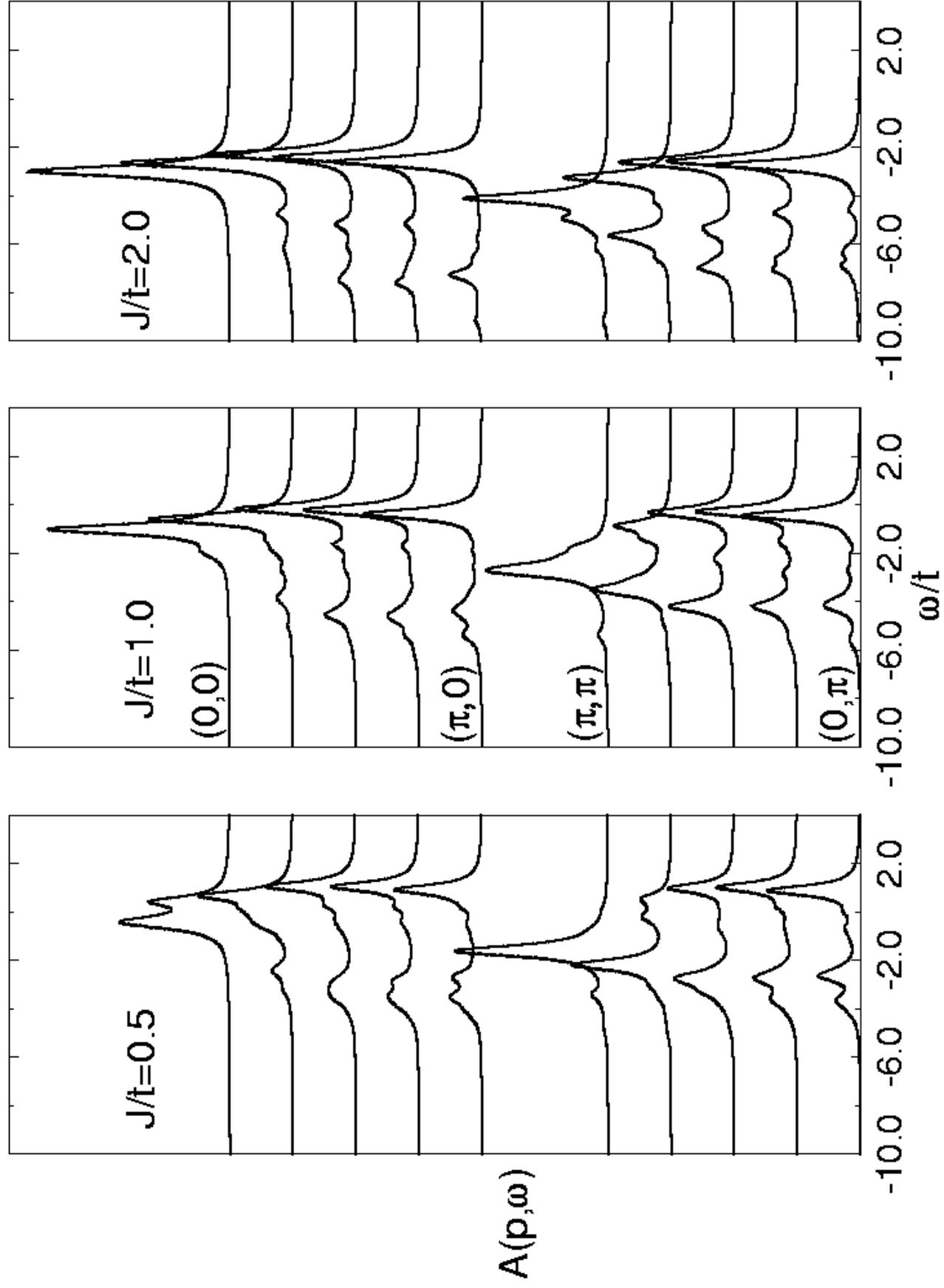
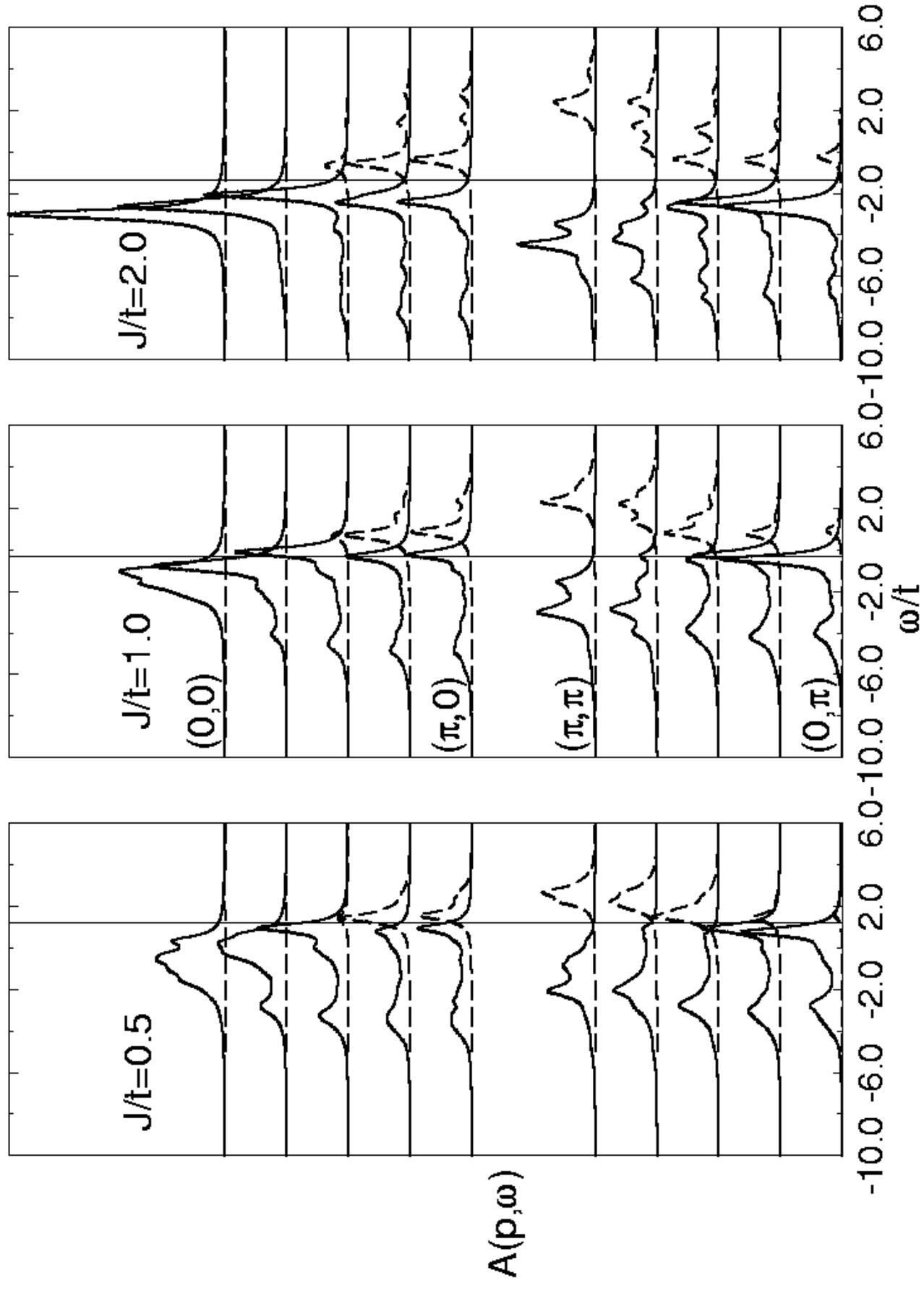
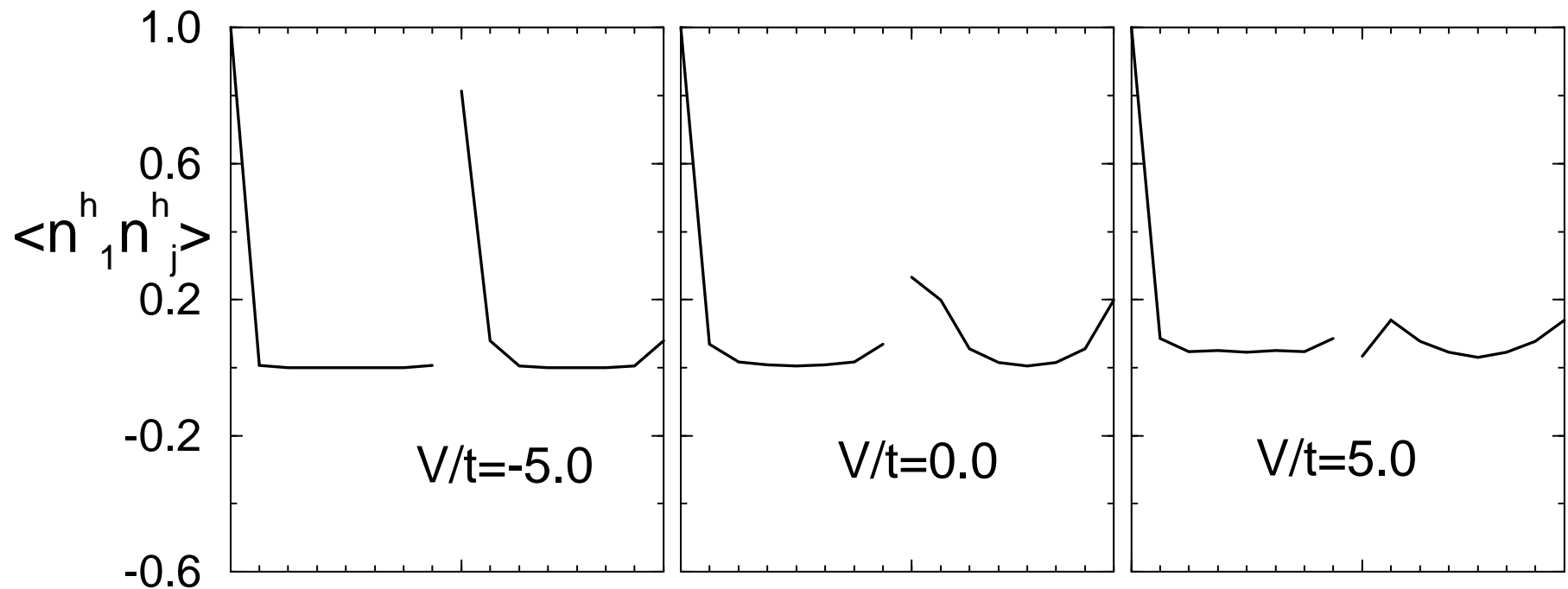
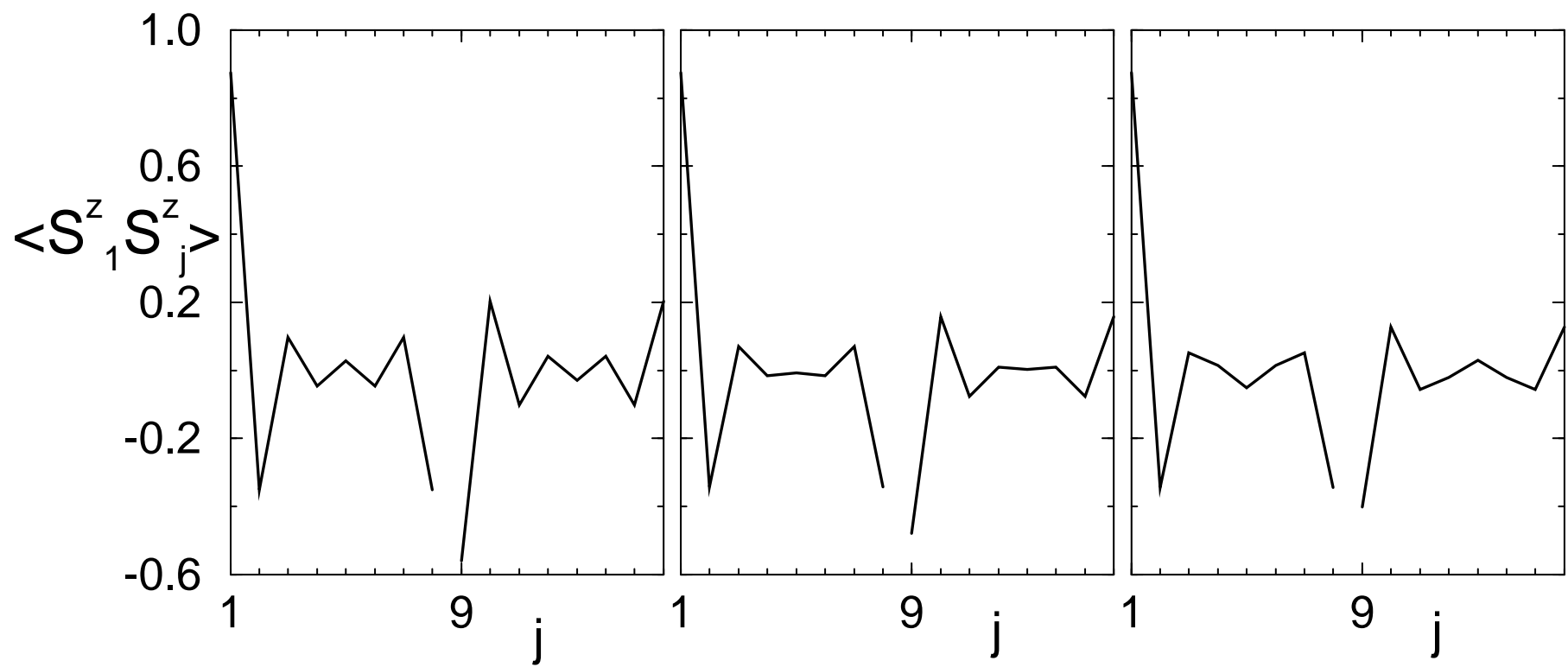
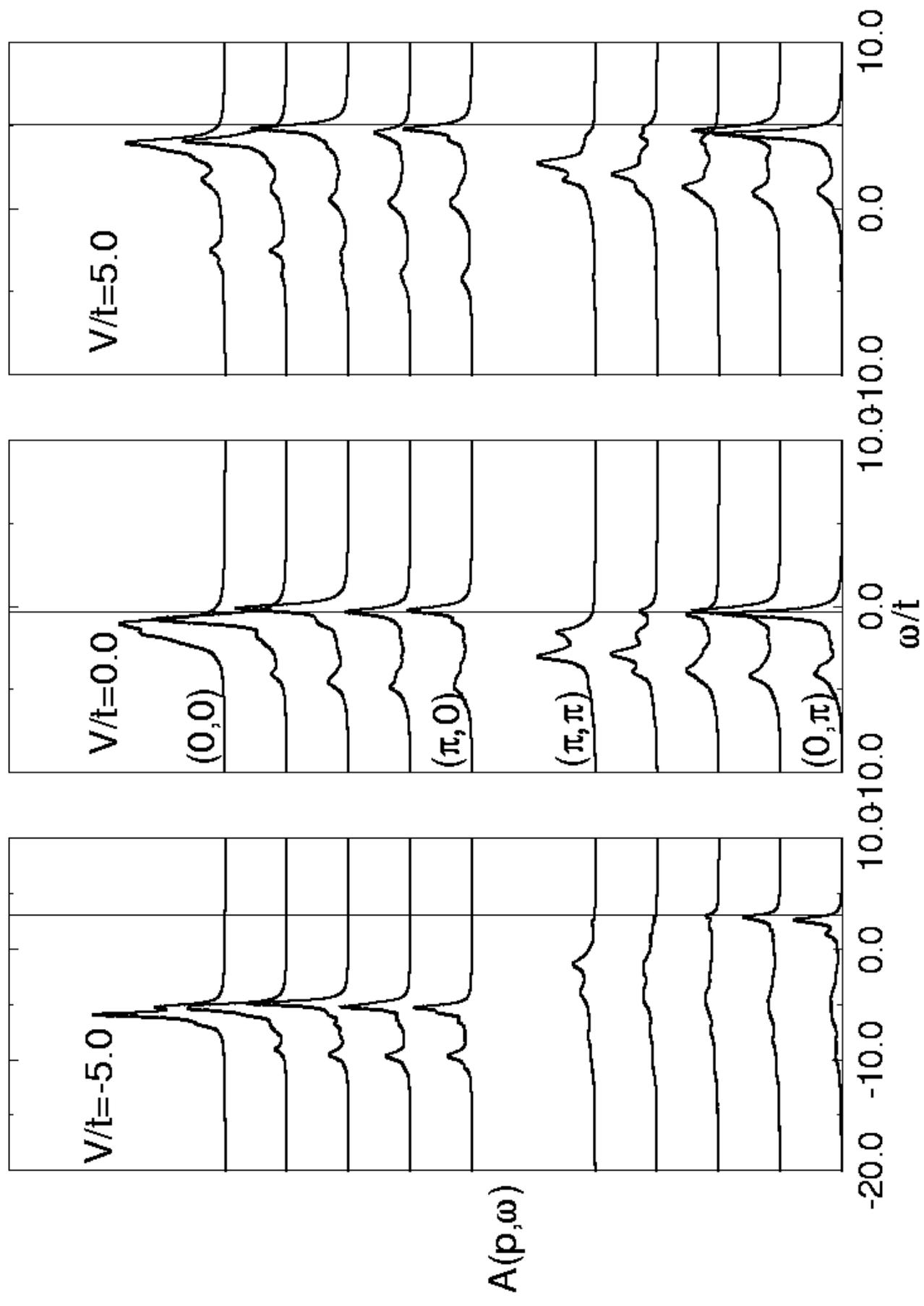
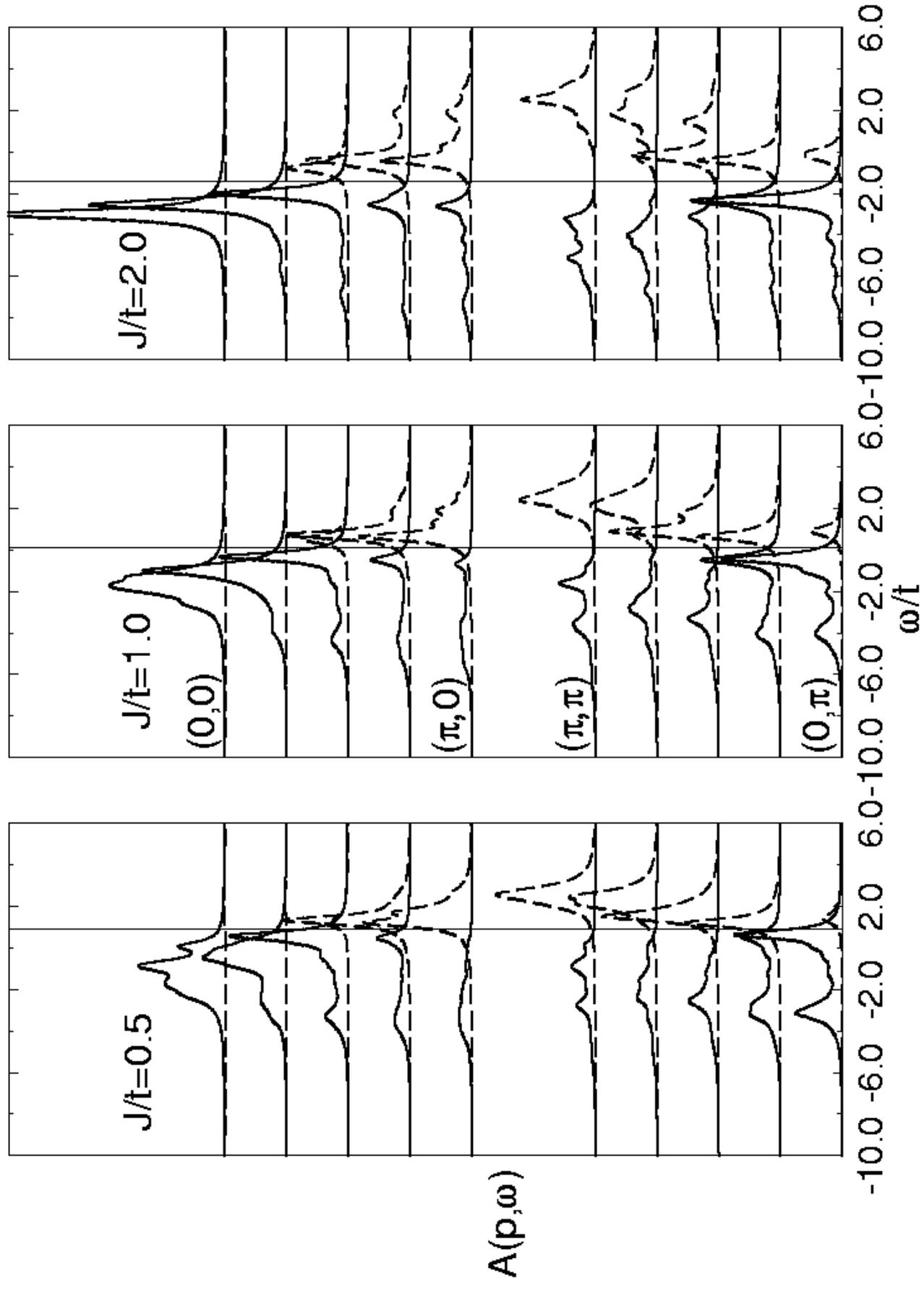
\end{document}